\documentclass{article}\usepackage[]{graphicx}\usepackage[]{xcolor}
\makeatletter
\def\maxwidth{ %
  \ifdim\Gin@nat@width>\linewidth
    \linewidth
  \else
    \Gin@nat@width
  \fi
}
\makeatother

\definecolor{fgcolor}{rgb}{0.345, 0.345, 0.345}

\usepackage{framed}
\makeatletter
 {\par\unskip\endMakeFramed%
 \at@end@of@kframe}
\makeatother

\definecolor{shadecolor}{rgb}{.97, .97, .97}
\definecolor{messagecolor}{rgb}{0, 0, 0}
\definecolor{warningcolor}{rgb}{1, 0, 1}
\definecolor{errorcolor}{rgb}{1, 0, 0}
\newenvironment{knitrout}{}{} % an empty environment to be redefined in TeX

\usepackage{alltt}

\usepackage[letterpaper,top=2cm,bottom=2cm,left=3cm,right=3cm,marginparwidth=1.75cm]{geometry}
\usepackage[T1]{fontenc}
\usepackage{appendix}
\usepackage{authblk}
\usepackage{authblk}
\usepackage{amsmath, amsfonts, amssymb}
\usepackage{graphicx}
\usepackage{algorithm}
\usepackage{amsthm}
\usepackage{bm}
\usepackage{cases}
\usepackage{caption}
\usepackage{xr-hyper}
\usepackage{hyperref}
\usepackage{xurl}

\DeclareMathOperator*{\argmin}{argmin}

\DeclareMathOperator{\Ex}{\mathbb{E}}

\usepackage{setspace}
\usepackage{parskip}

\usepackage{soul}
\usepackage{xcolor}

\usepackage{natbib}
\author{Aaron Gerding\thanks{Corresponding Author: \texttt{agerding@umass.edu}}}
\author{Nicholas G. Reich}
\author{Benjamin Rogers}
\author{Evan L. Ray}
\affil{Department of Biostatistics and Epidemiology, School of Public Health and Health
Sciences, University of Massachusetts at Amherst}

\title{Evaluating infectious disease forecasts \\ with allocation scoring rules}
\IfFileExists{upquote.sty}{\usepackage{upquote}}{}
\begin{document}

\newcommand{\del}[2]{\frac{\partial {#1} }{\partial {#2}} }
\newcommand{\dby}[2]{\frac{d {#1} }{d {#2}} }
\newcommand{\sbar}{\overline{s}}
\newtheorem{proposition}{Proposition}

\theoremstyle{remark}
\newtheorem*{remark}{Remark}

\maketitle

\begin{abstract}
Recent years have seen increasing efforts to forecast infectious disease burdens, with a primary goal being to help
public health workers make informed policy decisions. However, there has only been limited discussion of how
predominant forecast evaluation metrics might indicate the success of policies based in part on those forecasts. We
explore one possible tether between forecasts and policy: the allocation of limited medical resources so as to minimize
unmet need. We use probabilistic forecasts of disease burden in each of several regions to determine optimal resource
allocations, and then we score forecasts according to how much unmet need their associated allocations would have
allowed. We illustrate with forecasts of COVID-19 hospitalizations in the US, and we find that the forecast skill
ranking given by this allocation scoring rule can vary substantially from the ranking given by the weighted interval
score. We see this as evidence that the allocation scoring rule detects forecast value that is missed by traditional
accuracy measures and that the general strategy of designing scoring rules that are directly linked to policy
performance is a promising direction for epidemic forecast evaluation.
\end{abstract}


\section{Introduction}

Infectious disease forecasting models have emerged as important tools in public health outbreak response. The
predictions they provide increasingly inform decisions regarding a wide variety of countermeasures intended to reduce
transmission and mitigate the severity of disease outcomes. For example, estimates of the onset time of the flu season
have been used in developing national vaccination strategies \citep{igboh2023timing}, and forecasts of Ebola and
diptheria dynamics have been made with the clearly stated goal of helping local public health workers choose the timing
and location of interventions in settings where resources are severely constrained 
\citep{meltzer2014estimating, rainisch2015regional, camacho2015-ebola-bed,finger_real-time_2019}. More recently, in the
context of the outbreak of COVID-19 across the US, \cite{bertsimas2021predictionsCOVID} used forecasts as inputs to
decision tools for the interstate reallocation of ventilators and ICU capacity, and to recommend vaccine trial sites to
a major trial sponsor. \cite{fox_real-time_2022} similarly used predictive models to inform intrastate resource and
care site planning, as well as local community guidelines for masking, traveling, dining and shopping (University of
Texas, 2022)\nocite{utnews2022}.

In the wake of the COVID-19 pandemic, this trend has been followed by calls for infectious disease forecasts to be not
only designed, but also evaluated in ways that align specifically with how forecasts can be used to inform such outbreak
control decisions \citep{marshall2023predictions, bilinski_adaptive_2023}. This contrasts, however, with the
historically standard practice of measuring the quality of disease forecasts using general purpose accuracy and skill
scores, especially those that have implementations available in existing software when the relevant outbreak occurs. For
point forecasts, the root mean square error (RMSE) (e.g., \cite{papastefanopoulos2020covid}) and the mean absolute error
(MAE) (e.g., \cite{johansson2016evaluating}) are common choices. For probabilistic forecasts, which are the focus of
this paper, researchers have often relied on the logarithmic score (LS). For example, the LS has been used to evaluate
the skill of US seasonal influenza forecasts \citep{mcgowan_collaborative_2019,reich_collaborative_2019} as well as
forecasts targeting surveillance measures of dengue incidence in Peru and Puerto Rico \citep{johansson_open_2019}. More
recently, the continuous ranked probability score (CRPS) and a discretized version adapted to multi-quantile forecasts,
the weighted interval score (WIS) \citep{bracher2021evaluating}, have gained prominence. For example, the CRPS was used
to assess probabilistic forecasts (based on random effect models) of dengue incidence at the district level in Vietnam
\citep{colon-gonzalez_probabilistic_2021}. And the WIS has been used during the COVID-19 pandemic to evaluate forecasts
of observed cases, hospitalizations and deaths in the US and Europe, as reported by municipal, state, and federal
surveillance systems \citep{cramer_evaluation_2022,fox_real-time_2022,sherratt2023predictive}.

While it should be noted that there are ways to interpret any of these scores abstractly through the lens of decision
theory, and that all of the application-specific papers cited above benefited from direct collaboration with public
health agencies, a key impetus for the present work has been that we were not able to find in any of them, nor in the
literature they represent, explicit connections between how a forecast was evaluated and how that forecast was used in
practice.

A general phenomenon at play here --- one that has been observed repeatedly over the past few decades in other fields
such as finance, supply chain management, and meteorology --- is that while scores such as RMSE, MAE, LS, CRPS, and WIS
can describe the \emph{quality} of a forecast in terms of how well it corresponds to the observed disease outcome, they
will often fail to register the \emph{value} of a forecast in the context of a specific decision. We refer the reader to
\cite{yardley2021utility_cost_forecasts} and the references collected therein (especially the foundational
\cite{murphy1993whatisagoodforecast}) for a general overview touching on a wide range of forecasting contexts and  also
to \cite{pesaran2002decision_based_eval} for a clear discussion from an econometric perspective.

Despite this now well-developed discussion of the quality-value distinction in the larger forecasting community, we are
aware of only a limited literature attempting to connect the value of \emph{infectious disease} forecasts to their
impact on and through policy.  And within this body of work we have found the discussion of such a connection to still
be at a formally and quantitatively imprecise stage. In \cite{ioannidis2022forecastingCOVIDfailed}, the possible negative
consequences of inaccurate forecasts of infectious disease are discussed, but there is no attempt to quantify the
utility or loss incurred as a result of those forecasts. \cite{bilinski_adaptive_2023} explore ways in which predictive
classifiers of local COVID-19 risk levels in the US could be tuned to policymaker preferences for different costs
associated with over- and under-reaction to disease dynamics, but they do not clearly identify the source of these
costs or how they depend on quantifiable policy choices. A similar discussion related to dengue countermeasures in
Vietnam appears in \cite{colon-gonzalez_probabilistic_2021}.  A novel version of the WIS informally motivated by
utility considerations is developed in \cite{marshall2023predictions}, but the score is not derived in a
decision-theoretic manner.  There is also a thread of literature that frames infectious disease modeling as a component
of a larger system for understanding how policy goals, means, and choices interact and constrain one another. As an
example, \cite{Probert2016decisionMakingFootMouth} explore how policy recommendations ought to flow from a possibly 
incongruous set of simulation-based projection models of a hypothetical foot-and-mouth disease outbreak when there are 
ranges of plausible responses and stakeholder interests. Decision theory plays a prominent role here, but not explicitly
as a way to evaluate the choices made in developing the models.

In this work, we begin to fill this gap between the ways that infectious disease forecasts have traditionally been
evaluated and the ways that they have been used to support public health policy. To do so, we consider a setting in
which forecasts are used to help determine the allocation of a limited quantity of medical supplies across multiple
regions. In section~\ref{sec:methods} of the paper, we define a new forecast scoring rule --- the {\em allocation score}
--- that evaluates forecasts based on how beneficial resource allocations derived from them would turn out to be. In
section \ref{sec:application}, we present an illustrative analysis using the allocation score to evaluate forecasts of
hospital admissions in the US that were made leading up to and during the Omicron wave that peaked in January of 2022.
This analysis is ``synthetic'' insofar as it is not intended to correspond to any specific historical record of
allocation decisions that could have been supported by hospitalization forecasts during this period. However, we view
the general allocation problem on which our framework is based as a versatile template for formalizing real-world
decisions that must constantly be made in real-time by public health administrators around the globe, especially those
related to hospital capacity, ventilator usage, doses needed for specific medications and other situations where an
outbreak creates sudden and highly variable demand for potentially scarce resources.  

\section{The Allocation Score}
\label{sec:methods}

We begin with an informal description of the allocation score and some examples illustrating its key characteristics in
section \ref{sec:methods.overview}. In section \ref{sec:methods.detailed} we develop the allocation score more
carefully, using a decision theoretic procedure for deriving proper scoring rules. (See appendix \ref{sec:a:proper}
for a definition of a proper scoring rule and a more technical discussion of the procedure). In section
\ref{sec:methods.related}, we note that another group of common scores including the quantile score, WIS, and CRPS, can
also be derived from decision theoretic foundations \textemdash starting from a different decision making context.

\subsection{Overview of Allocation Scoring}
\label{sec:methods.overview}

Suppose that a decision maker is tasked with determining how to allocate $K$ available units of a resource across $N$
locations. If the decision maker is provided with a multivariate forecast $F$ where each marginal forecast distribution
$F_i$ predicts resource need in a particular location, one option is to choose the resource allocation that minimizes
the expected total unmet need according to the forecast. We will give a more precise mathematical statement in section
\ref{sec:methods.detailed}, but informally, the total expected unmet need according to the forecast is
\begin{align}
\sum_{i=1}^N \mathbb{E}_{F_i}[\text{unmet need in location $i$}], \label{eqn:informal_objective}
\end{align}
where the unmet need in a particular location is the difference between resource need in that location and the number of
resource units that were allocated there. This allocation problem has an intuitively appealing solution: allocate so that the
probabilities of need exceeding allocation in various locations are as close to each other as possible. This will lead
to the allocations provided by $F$ being quantiles of the marginal distributions $F_i$ for some \emph{single}
probability level $\tau$ that is shared in common for all locations.

After time passes and the actual level of resource need has been observed, the value of a selected allocation can be
measured by comparing the actual need in each location to the amount of resources that were sent there. Specifically, we
compute the total unmet need that resulted from the selected allocation:
\begin{align}
    \sum_{i=1}^N \text{unmet need in location $i$}. \label{eqn:informal_loss}
\end{align}
We regard one allocation as better than another (with respect to the realized need) if it results in lower total 
unmet need, and accordingly regard one forecast as better than another (again with respect to the realized need)
if the allocation derived from the former results in lower total unmet need than does the allocation derived from the 
latter.

The \textbf{allocation score} of the forecast $F$ is the avoidable unmet need that results from using the allocation
that minimizes the expected unmet need according to that forecast. By ``avoidable unmet need'', we mean that the
allocation score does not include the amount of unmet need that was inevitable simply because the amount of available
resources $K$ was less than the need for resources. Rather, the allocation score measures the unmet need that could have
been avoided by an oracle that knows exactly how much need will occur in each location and divides the amount $K$ so
that nothing is wasted in one location while it could be put to use in another. The best and lowest possible allocation 
score is 0.  A positive allocation score indicates that an oracle with perfect foreknowledge could have selected an allocation 
that met more need than the allocation suggested by $F$.

\begin{figure}
    \includegraphics[width=\textwidth]{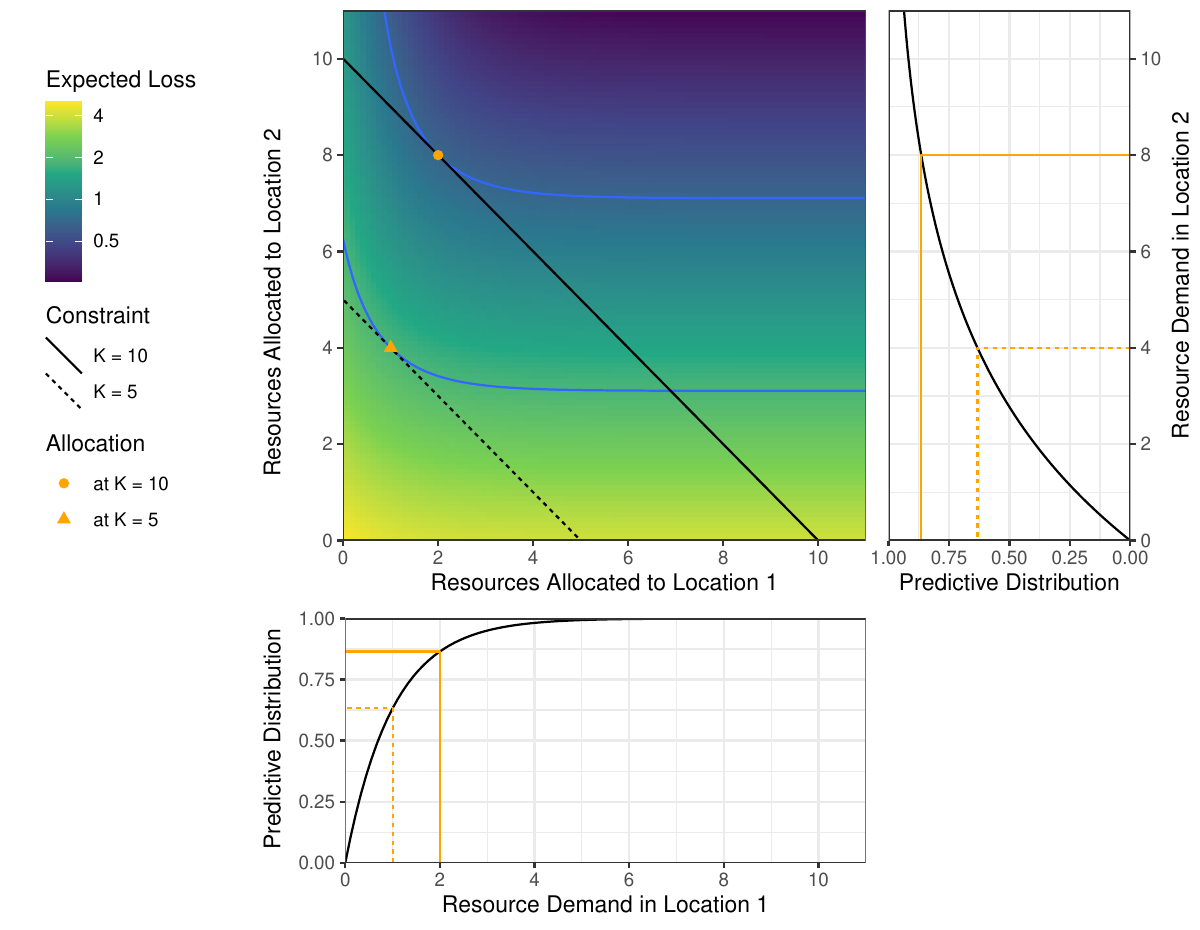}
    \caption{An illustration of the resource allocation problem in Example 1. There are $N = 2$ locations, with
    predictive distributions $F_1 = \mathrm{Exp}(1)$ and $F_2 = \mathrm{Exp}(1/4)$. The cumulative distribution
    functions of these distributions are illustrated in the panels at bottom and right. In the center panel, the
    background shading corresponds to the expected loss according to these forecasts. Diagonal black lines indicate
    resource constraints at $K=5$ and $K=10$ units; any point along those lines corresponds to an allocation that meets
    the resource constraint. For these forecasts, the optimal allocations are $(1, 4)$ for $K=5$ and $(2, 8)$ for
    $K=10$. These allocations are at the point on the constraint line where the expected loss is smallest, which also
    corresponds to the point where a level set of the expected loss surface (blue curve) is tangent to the constraint.}
    \label{fig:exp_alloc_example}
\end{figure}

\paragraph{Example 1} Suppose we have a forecast $F$ for need in two locations such that $F_1$ and $F_2$ have exponential marginal distribution with scale parameters $\sigma_1 = 1$ and $\sigma_2 = 4$. The means of these distributions are
given by the scale parameters $\sigma_i$. When the marginal forecasts are exponential distributions, it can be shown
that the optimal allocation divides the available resources among the locations proportionally to the scale parameters
$\sigma_i$ (see appendix \ref{sec:a:bayes-quantiles}). If $K = 5$ units of the resource are
available, the optimal allocation according to $F$ would be 1 and 4 resource units in locations 1 and 2, respectively. 
Increasing $K$ to $10$ changes this optimal allocation to 2 and 8. Figure~\ref{fig:exp_alloc_example} illustrates the 
situation.

Next suppose that we observe resource needs of 1 and 10 in locations 1 and 2, respectively. Based on these observed
needs, we evaluate the allocation suggested by the forecast by calculating the amount of unmet need that resulted from
that allocation over and above what was unavoidable given the resource constraint. With $K = 5$ available resource
units, the allocation based on the forecast exactly meets the observed need in location 1, but it leaves 6 units of need
unmet in location 2. But these 6 units of unmet need are unavoidable under the constraint $K=5$: allocating 0 units of
resources to location 1 and 5 to location 2, for example, still results in a total unmet need of 6 across both
locations. This gives an allocation score of 0 for the forecast when $K = 5$. On the other hand, when $K = 10$, the
forecast's allocation results in $10 - 8 = 2$ units of unmet need in location 2 despite leaving no need unmet in
location 1. In this case, the oracle would be able to prevent all but 1 of the total 11 units of need from going unmet,
for example by allocating 1 unit of resources to location 1 and the remaining 9 units of resources to location 2. So for
$K = 10$, the forecast gets an allocaton score of 1 (= 2 realized $-$ 1 unavoidable) in units of avoidable unmet need.

These scores illustrate a general result: allocation scores for a forecast will tend to be larger when the resource
constraint is close to the observed need, because this is when it matters most which locations are allocated more or
less resources. If the amount of available resources is small relative to the observed need, any allocation of those
limited resources will result in a large amount of unmet need. If the amount of available resources is comparatively
large, it becomes less important which locations receive relatively more or fewer resources because all locations are
likely to
receive enough resources to meet their need. In either of these extremes of resource availability, the avoidable unmet
need that arises from the allocation suggested by a forecast (i.e., the forecast's allocation score) will tend to be
small.

\paragraph{Example 2} Now consider a different forecast that also has exponential distributions for resource need in
each location, but that has the scale parameters $\sigma_1 = 2$ and $\sigma_2 = 8$, twice as large as the scale
parameters of the forecast in Example 1. Because the optimal allocation is proportional to the scale parameters, this
forecast would lead to the same allocations as the forecast in Example 1, and would therefore be assigned the same
allocation score.

Examination of these results leads to two observations. First, the reason that these forecasts had a positive (i.e.,
non-optimal) allocation score at $K=10$ is that they did not get the relative magnitude of resource need across the two
locations right: the realized need was 10 times as large in location 2 as in location 1, but the forecasts only
indicated that the resource allocation for location 2 should be 4 times the allocation for location 1. At its core, the
allocation score measures whether the forecast accurately captures the relative magnitudes of resource need across
different locations, which is precisely the information that is needed to allocate resources to those locations subject
to a fixed resource constraint.

A second observation is that the forecasts in examples 1 and 2 predicted different mean levels of resource needs, but
had the same allocation score. The allocation score does not directly measure whether the forecasts correctly capture
the absolute magnitude of resource need in each individual location. This stands in contrast to other common scoring
methods that aggregate scores such as log score, CRPS, or WIS for each location, where a poor forecast for one location
is penalized regardless of alignments in other units.

\subsection{A decision theoretic development of the allocation score}
\label{sec:methods.detailed}

We give a high-level review of a general procedure for developing proper scoring rules that are tailored to specific
decision making tasks in section \ref{sec:methods.detailed.decisiontheory}, and then in section
\ref{sec:methods.detailed.specific_allocation} we apply that procedure to develop the allocation score based on the task
of deciding on how to allocate a fixed supply of resources across multiple locations. In
\ref{sec:methods.detailed.integrated_allocation} we consider a small extension where the resource constraint is not
known, or it is desired to consider the value of forecasts across a range of decision making scenarios. This gives rise
to the \emph{integrated allocation score}.

\subsubsection{The decision theoretic setup for forecast evaluation}
\label{sec:methods.detailed.decisiontheory}

Decision theory investigates how decisions are made by formalizing a decision as the process of selecting an
\emph{action} $x$ from a specified set $\mathcal{X}$ of potential actions while taking into consideration the possible
consequences of $x$ under future states of the world. Let $Y$ be a quantifiable aspect of the future which will help to
determine the consequences of $x$. $Y$ is a random variable inasmuch as it is indeterminate when the decision is made.
We refer to a realization $y$ of $Y$ as an \emph{outcome}. A fundamental strategy in decision theory is to assume that
the consequences of an action $x$ under outcome $y$ can be assigned a numeric value, or \emph{loss}, measuring the (lack
of) success of $x$.

Some examples of actions are levels of investment in measures designed to mitigate severe disease outcomes such as
hospital beds, ventilators, medication, or medical staff, with $\mathcal{X}$ being the set of all feasible levels of
investment. An outcome $y$ determining the consequences of such an action might be the number of individuals who
eventually become sick and would benefit from the mitigation measure. In our example, $x$ will succeed to the extent
that it meets the realized need $y$, so that greater unmet need will incur greater loss.

When a forecast $F$ of $Y$ is available, a decision maker might choose that action $x$ which yields the lowest expected
loss according to $F$. The loss of the action $x$ chosen according to $F$ can then, in turn, be interpreted as the value
of $F$ as an input to the decision making process under the realized outcome $y$ of $Y$.

The strategy of minimizing the expected loss according to $F$ might be used if the decision maker trusted that $F$ was
accurate and did not have any additional information that was not captured in $F$. However, here we use it only as a
device for evaluating forecasts by examining the consequences of hypothetical decisions that would be made when using
only $F$ as an input. We return to this point in the discussion.

We arrive at a three-step procedure for developing scoring rules for probabilistic forecasts:
\begin{enumerate}
  \item Specify a \emph{loss function} $s(x, y)$ that measures the loss associated with taking action $x$ when outcome $y$
    eventually occurs.
  \item Given a probabilistic forecast $F$, determine the \emph{Bayes act} $x^F$ that minimizes the expected loss under
    the distribution $F$.
  \item The \emph{scoring rule} for $F$ calculates the score as the loss incurred when the Bayes act was used: 
    $S(F, y) = s(x^F, y)$.
\end{enumerate}
This is a general procedure that may be applied in settings where it is possible to specify a quantitative loss
function. We call a scoring rule obtained from this procedure a \emph{Bayes scoring rule} (noting that to our knowledge, 
this terminology is not standard).  In appendix \ref{sec:a:proper} we review the
result from the scoring rule literature (e.g., \cite{dawid2007geometry,gneiting2007strictly}) that Bayes scoring
rules are proper by construction.

\subsubsection{The allocation score for a fixed resource constraint}
\label{sec:methods.detailed.specific_allocation}

In our setting an action is a vector $x = (x_1, \ldots, x_N)$ specifying the amount of resources allocated to each of
$N$ locations. We require that each $x_i$ is non-negative, and that the total allocation across
all locations equals the amount of available resources, $K$: $\sum_{i=1}^N x_i = K$. The set $\mathcal{X}$ consists of
all possible allocations that satisfy these constraints. The eventually realized resource need in each location is
denoted by $y = (y_1, \ldots, y_N)$; this is a realization (unknown at the
time of decision making) of the random vector $Y = (Y_1, \ldots, Y_N)$. A forecast of the random need $Y$ for each
location is written as $F = (F_1, \ldots, F_N)$. We assume need is non-negative and that the forecasts reflect that, i.e. the
support of each $F_i$ is a subset of $\mathbb{R}^+$. Finally, we assume that each unmet unit of need in any location
incurs a fixed loss of $L>0$, so that if the selected resource level $x_i$ in location $i$ is less than the realized
need $y_i$, a loss of $L \cdot (y_i - x_i)$ results. A variety of extensions to this setup are possible; for example, we
might account for storage costs for resources that go unused, allow for a different loss per unit of unmet need in each
location, or account for resource transportation costs. In this work, we choose to keep the loss function relatively
simple to focus on the core ideas.

It is helpful to clearly distinguish between the time $t_d$ when a decision is made about a public health
resource allocation and the time $t_r$ when resource needs that might be addressed by that allocation occur. Our
setup assumes that $t_d < t_r$ and that the resource in question can only meet the need realized at $t_r$, not prevent
it. That is, we stipulate that the choice of allocation $x$ has no effect on the distribution of the random vector $Y$. In the
context of infectious disease, this means that we do not consider resources, such as vaccines, that are intended to
reduce the number of people who will become sick at some point in the future. Instead, our setup addresses resources
like hospital beds, oxygen supply, or ventilators which are intended to meet the medical needs of patients who are
already sick. Additionally, our setup addresses decision-making that is related to resource needs only at the time
$t_r$; we do not consider sequences of multiple decisions that are made over time or account for the impact of decisions
on resource needs at any time other than $t_r$. We outline some opportunities to extend our work to more complex
decision making settings in the discussion.

With this problem formulation in place, we can develop a proper scoring rule following the procedure in section
\ref{sec:methods.detailed.decisiontheory}.

\paragraph{Step 1: specify a loss function.} The loss incurred by an allocation $x$ under an outcome $y$ is the sum of
contributions from unmet need in each location:
\begin{equation}
  s_A(x, y) = \sum_{i=1}^N L \cdot \max(0, y_i - x_i). \label{eqn:loss_fn}
\end{equation}
Here, $\max(0, y_i - x_i)$ is the unmet need in location $i$, which is given by $y_i - x_i$ if the realized need $y_i$
in location $i$ is greater than the amount $x_i$ allocated to that location, or $0$ if the amount $x_i$ allocated to
unit $i$ is greater than or equal to the realized need. Also, $L$ is a constant scalar value, the same across all
locations, specifying the ``cost'' of one unit of unmet need.

\paragraph{Step 2: Given a probabilistic forecast $F$, identify the Bayes act.} The Bayes act associated with the
forecast, $x^{F,K}$, is the allocation that minimizes the expected loss, that is, the solution of the \emph{allocation
problem} associated with $K$:
\begin{align}
  \underset{0 \leq x}{\mathrm{minimize}}\,\, \mathbb{E}_{F} [s_A(x, Y)] \text{ subject to }
  \, \sum_{i=1}^N x_i = K, \label{eqn:AP}
\end{align}
where $\mathbb{E}_{F} [s_A(x, Y)] = \sum_{i=1}^{N} L \cdot \mathbb{E}_{F_i}[\max(0, Y_i - x_i)]$ sums the expected loss
due to unmet need across all locations.

In appendix \ref{sec:a:bayes-quantiles} we show that the components of the Bayes act are quantiles 
$x_i^{F,K} = F_i^{-1}(\tau^{F,K})$ at a probability level $\tau^{F,K}$ that depends on the forecast $F$ and the resource
constraint $K$, but is shared across all locations. This probability level is the level at which the resource constraint
is satisfied: $\sum_{i=1}^N F_i^{-1}(\tau^{F,K}) = K$. This tells us that in order to allocate optimally (according to
$F$), we must divide resources among the locations so that there is an equal forecasted probability in every location
that the allocation is sufficient to meet resource need. This solution to the allocation problem is well-known in
inventory management and is often attributed to \cite{hadleywhitin1963}.

\paragraph{Step 3: Define the scoring rule.} We can now use the Bayes act to define a proper scoring rule for the
probabilistic forecast $F$. Consider first the raw score defined as
\begin{align}
  S_A^{\text{raw}}(F, y; K) = s_A(x^{F,K}, y) = \sum_{i=1}^N L \cdot \max(0, y_i - x_i^{F,K}).
\end{align}
This measures the total unmet need across all locations that results from using the Bayes allocation associated with the
forecast $F$ when the actual level of need in each location is observed to be $y_i$.

To make this a more easily interpreted measure of forecast performance, we will adjust the raw score by subtracting the
minimum loss achievable by an \emph{oracle} allocator which has precise foreknowledge of the outcomes $y_i$. When the
oracle has sufficient resources to meet the total need, i.e., when $\sum_{i=1}^{N}y_i \leq K$, the oracle's loss is zero
and allocation score coincides with the raw score. On the other hand, when the oracle cannot cover all need and incurs a
loss of $L \cdot \left(\sum_{i=1}^{N}y_i - K \right) > 0$, we adjust the raw score by this loss.
The oracle-adjusted score can therefore be written as
\begin{align}
  S_A(F, y; K)  &= S_A^{\text{raw}}(F, y; K) - L \cdot \max\left(0,\sum_{i=1}^{N}y_i - K\right) \\
  &= L\left\{\sum_{i=1}^N \max(0, y_i - x_i^{F,K}) -  \max\left(0,\sum_{i=1}^{N}y_i - K\right)\right\}.
\end{align}
The oracle adjustment aligns with a common theme in economic decision theory that opportunity loss (often known
as regret or (negative) relative utility) is often a more important quantity than absolute loss (see e.g.,
\cite{DIECIDUE201788}).

\subsubsection{Integrating the allocation score across resource constraint levels}{}
\label{sec:methods.detailed.integrated_allocation}

The allocation score $S_A$ developed in the previous section evaluates the forecast distributions
$F$ based on a single probability level $\tau^{F,K}$. This is appropriate if the resource constraint $K$ is a known
constant. However, if $K$ is not precisely known at the time of decision making or there is interest in measuring the
value of forecasts across a range of decision making scenarios with different resource constraints, we can use an
\emph{integrated allocation score} (IAS) that integrates the allocation score across values of $K$, weighting by a
distribution $p$:
$$S_{IAS}(F, y) = \int S_A(F,y; K) p(K) \, dK$$
We note that the device of considering a range of hypothetical decision makers or decision making problems with
different problem parameters has been employed in the past \cite[e.g.,][]{murphy1993whatisagoodforecast}.

\subsection{Connections to Other Scores}
\label{sec:methods.related}

The weighted interval score (WIS) was proposed in 2020 as a way to score forecasts that were being made in the early
stages of the COVID-19 pandemic \citep{bracher2021evaluating}; equivalent scores had also been used in previous forecast
evaluation efforts \cite[e.g.,][]{hong2016probabilisticEnergyForecasting}. The WIS is a proper scoring rule for
forecasts that use a set of quantiles to represent a probabilistic forecast distribution. Scores are calculated as a weighted sum of
interval scores at different probability levels (e.g., 50\% prediction intervals, 80\% PIs, 95\% PIs, etc...). Larger
interval scores indicate less skillful forecasts. An interval score consists of (a) the width of the interval, with
larger intervals receiving higher scores (higher scores indicate less accuracy), and (b) a penalty if the interval does
not cover the eventual observation, which increases the further away the interval is from the observed value.
Equivalently, the WIS can also be characterized as a weighted sum of quantile scores for each individual predictive
quantile. The quantile score for a particular quantile level assigns an asymmetric penalty to predictions that are too
high or too low, with the relative sizes of the penalties set so that in expectation the score is minimized by the given
quantile of the distribution. The most commonly used version of WIS is one that uses an equal weighting of all quantile
levels, in which case WIS approximates the continuous ranked probability score (CRPS), a commonly used score for
probabilistic forecasts. It is important to note that this weighting was proposed because the resulting score
approximates the CRPS, and not because it aligned with any particular public health decision-making rationale. For more mathematical detail on the WIS, we point readers to \cite{bracher2021evaluating}.

That said, the quantile score and WIS can be derived using the same decision theoretic procedure that we outlined in
section \ref{sec:methods.detailed}. In fields such as meteorology and supply chain management, a great deal of attention
has been given to the problem where a decision must be made about the quantity of a resource to purchase for a single
location in the face of a fixed cost $C$ for each unit of the resource and a loss $L$ that will be incurred for each
unit of unmet need. This leads to the quantile score for the probability level $\tau = 1 - C/L$. From this point, the
WIS or CRPS can be obtained by averaging across a range of decision making settings with different cost and loss
parameters, using a similar motivation that we used to obtain the IAS from the AS in section
\ref{sec:methods.detailed.integrated_allocation} \citep{gneiting2011weightedScoringRules}.

\section{Evaluating forecasts of COVID hospitalizations using the allocation score}
\label{sec:application}

We illustrate our new forecast evaluation framework with an application to hospital admissions in the U.S., considering
a hypothetical problem of allocating a limited supply of medical resources to states.

\subsection{Data}

The US COVID-19 Forecast Hub collected short-term forecasts of daily new hospital admissions for individuals with
COVID-19 starting in December 2020 \citep{cramer_united_2022}. The target data for these forecasts were hospital
admissions as reported by the US Department of Health and Human Services through the HealthData.gov website. Forecasts
were probabilistic predictions of the number of new hospital admissions on a particular day in the future, in a specific
jurisdiction of the US (national level, state, or territory). Probability distributions were represented using a set of
23 quantiles for each individual prediction, including a median and the lower and upper limits of 11 central prediction
intervals, from a 99\% to a 10\% prediction interval.

The analysis in this work focuses on forecasts made before and during the first wave of the Omicron SARS-CoV-2 variant
in the US. As such, we analyzed forecasts for the 15 weeks starting with Monday November 22, 2021 through Monday
February 28, 2022.

Submission to the Forecast Hub followed a weekly cycle, and each Monday the Hub collected the most recent forecasts
submitted by all teams that met certain inclusion criteria and created ensemble forecasts using quantile averaging
\citep{ray_comparing_2023}. Our analysis includes these ensembles (\texttt{COVIDhub-ensemble} and
\texttt{COVIDhub-trained\_ensemble}) as well as one other ensemble of hub models created by another team
(\texttt{JHUAPL-SLPHospEns}) and several other individual models. Models were eligible to be included in the analysis if
they were designated as a ``primary'' model from a team. For a model to have a complete, eligible submission in a given
week, it had to have a 14 day-ahead forecast for all 50 states plus Washington DC. Models had to have a complete
forecast for at least 4 of the 15 weeks in the analysis to be eligible for inclusion.

The hospitalization data used for scoring forecasts were downloaded on February 07, 2024.

\subsection{Evaluation metrics}

We evaluated forecasts using the allocation score (AS) and the weighted interval score (WIS), computed at a horizon of
14 days. We chose to focus much of our analysis on the AS computed for a resource constraint of $K=15,000$. This level
provides an anchor for interesting comparisons between phases of the Omicron wave by representing a national shortage at
the peak and a national excess before and after the wave (see Figure \ref{fig:metrics-over-time}A), assuming that the
resource need corresponds directly to new hospital admissions. We additionally computed the mean WIS (MWIS) across all
of the forecasted state-level locations. 

For both scores, we also computed standardized ranks among all models that submitted forecasts each week. The
standardized rank lies in the interval $[0, 1]$, where 0 corresponds to the worst rank and 1 to the best. In the case of
a tie between one or more models, all models received the better rank.

As described above, predictions were submitted to the Forecast Hub in the form of a set of 23 quantiles of the
predictive distribution. The WIS can be directly calculated from these quantiles. However, our numerical method for calculating 
the AS (outlined in appendix \ref{sec:a:numeric})
requires a full cumulative distribution function (CDF) for the forecast in each location.
For the purpose of this analysis, we imputed CDFs based on the provided quantiles following a procedure detailed in 
appendix \ref{sec:a:distfromq}. Briefly, this involves interpolating the provided quantiles between the lowest (.01) and highest 
(.99) probability levels using monotonic cubic splines, and then extending outside these levels with normal distribution tails
parametrized to match the two lowest and two highest quantiles. We show in appendix \ref{sec:a:propriety_of_parametric_approximation}
that if this evaluation procedure
had been specified prospectively, the resulting score would be proper, but that a post hoc application of this
procedure is improper. We use the procedure here to illustrate the properties of the AS, and note that a collaborative
forecasting hub interested in using the AS for evaluation could circumvent propriety issues by communicating the
definition of the AS to forecasters and then collecting allocations at specified resource levels as part of forecast
submissions.

\subsection{Allocation benchmarks}

In order to explore how model-derived allocations might be compare to ``standard operating procedures'' for allocation
used by public health officials, we scored two simple benchmark methods. First, we evaluated forecasts from the
\texttt{COVIDhub-baseline} model, which predicts a flat line from the most recent observation with uncertainty bounds
based on a random walk \citep{cramer_evaluation_2022}. Second, we generated a proposed set of allocations where the
quantity allocated to each state was proportional to that state's population (using US Census data of vintage 2022
\citep{us_census_nst_est2022}), referred to below as \texttt{per-capita}. These two approaches generally reflect common
choices for ``best practices'' of allocation: either allocating resources based on the most recent observed data, or in
proportion to the population of each location. 

\subsection{Data and code availability}

All forecast data used in this evaluation are available through the COVID-19 Forecast Hub \citep{cramer_united_2022}. An
R package implementing the allocation score is available at \url{https://github.com/aaronger/alloscore}. All code and
data for the analyses presented in this manuscript are available at
\url{https://github.com/aaronger/utility-eval-papers}. The analyses were generated using a reproducible workflow using
R version 4.3.1 (2023-06-16) and the {\tt targets} package \citep{Rcore-2023, landau_2021_targets}.

\subsection{Application results}

\subsubsection{Anatomy of forecast scores for one week}

To illustrate the mechanics of allocation scoring, we start by focusing on how forecasts generated on or before December
20, 2021, with predictions for January 03, 2022, were scored by different metrics. This week was around the peak of the
Omicron wave nationally, with individual states typically observing a peak at or after January 3, 2022.

Of the 10 models evaluated for this one week, the CU-select model had the most accurate forecasts according to the
allocation score while the USC-SI\_kJalpha model had the most accurate forecasts based on MWIS (Table
\ref{tab:multi-k-scores}). The \texttt{JHUAPL-Bucky} model had the second best MWIS but the third worst allocation score.

% latex table generated in R 4.3.1 by xtable 1.8-4 package
% Fri Mar  1 16:18:50 2024
\begin{table}[ht]
\centering
\begin{tabular}{lrrrr}
  \hline
Model & AS & MWIS & IAS centered at 15k & IAS uniform \\ 
  \hline
\texttt{CU-select} & 669 & 133 & 774 & 326 \\ 
  \texttt{per-capita} & 865 & - & 1029 & 366 \\ 
  \texttt{COVIDhub-ensemble} & 873 & 159 & 1067 & 438 \\ 
  \texttt{USC-SI\_kJalpha} & 995 & 91 & 1216 & 1097 \\ 
  \texttt{JHUAPL-Gecko} & 1034 & 164 & 1141 & 418 \\ 
  \texttt{MUNI-ARIMA} & 1084 & 169 & 1248 & 440 \\ 
  \texttt{COVIDhub-trained\_ensemble} & 1089 & 169 & 1271 & 823 \\ 
  \texttt{COVIDhub-baseline} & 1175 & 170 & 1317 & 535 \\ 
  \texttt{JHUAPL-Bucky} & 1358 & 102 & 1566 & 1214 \\ 
  \texttt{JHUAPL-SLPHospEns} & 1540 & 129 & 1604 & 1102 \\ 
  \texttt{UVA-Ensemble} & 2469 & 213 & 2635 & 2494 \\ 
   \hline
\end{tabular}
\caption{For one illustrative week, a comparison of allocation scores (AS), mean weighted interval scores (MWIS), and two varieties of Integrated Allocation Scores (IAS), (see Section \ref{sec:ias_examp}). All metrics are shown for 10 models that made forecasts of hospital admissions for 2022-01-03. Results are sorted by AS. For all metrics, lower scores indicate better accuracy.} 
\label{tab:multi-k-scores}
\end{table}

\begin{knitrout}
\definecolor{shadecolor}{rgb}{0.969, 0.969, 0.969}\color{fgcolor}\begin{figure}[H]
\includegraphics[width=\maxwidth]{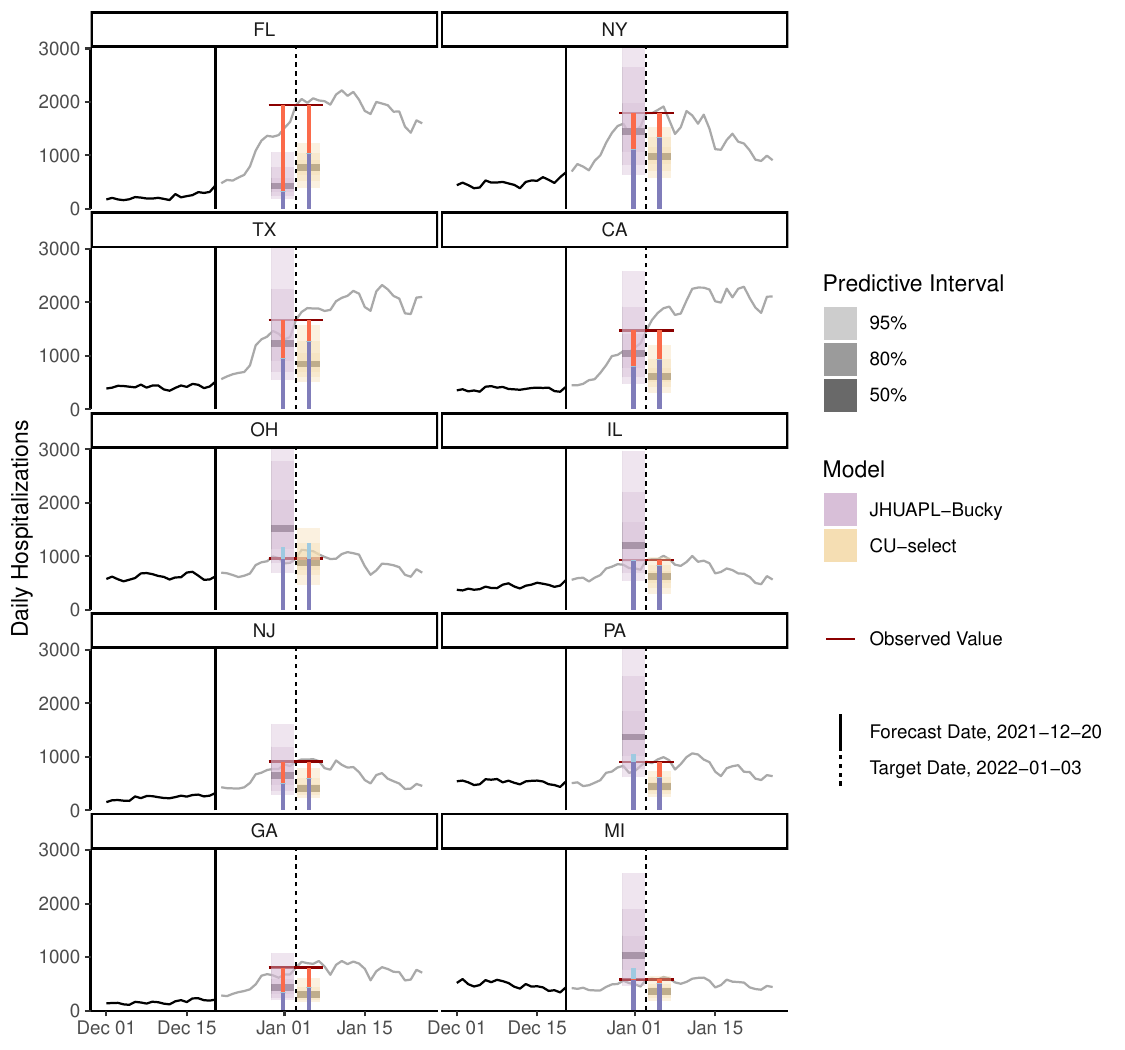} \caption[Probabilistic forecasts for new COVID-19 hospital admissions on January 3, 2022 and their suggested resource allocations for the states with the ten highest (eventually observed) hospitalization counts]{Probabilistic forecasts for new COVID-19 hospital admissions on January 3, 2022 and their suggested resource allocations for the states with the ten highest (eventually observed) hospitalization counts. For each state, the dark black line shows the data observed when the forecast was made, the grey line shows counts observed after the forecast was made, and the horizontal dark red segment shows the count observed on the target date. The side-by-side shaded regions show the median (grey horizontal line) and 50\%, 80\% and 95\% prediction intervals for the two selected models. The forecasts were made for new hospitalizations on January 3, 2022 (vertical dashed line, with number of hospitalizations indicated by red horizontal line segment at the intersection of the dashed line and the grey line of data). The vertical bars with purple, blue and red shading show the allocations. The purple bar goes from zero to the amount of need that was met by the allocation from that model to a specific location. A red bar indicates need that exceeded the resource allocation for a location, and a light blue bar shows the amount by which the resource allocation suggested by that model exceeded the need for that location.}\label{fig:thermometer-plot}
\end{figure}

\end{knitrout}

A state-wise comparison of the
\texttt{JHUAPL-Bucky} and \texttt{CU-select} models shows that while the \texttt{JHUAPL-Bucky} forecast distributions
were more centered on the eventual observations in many states, the suggested allocations of the \texttt{JHUAPL-Bucky}
model were less efficient than those of the \texttt{CU-select} model (Figures \ref{fig:thermometer-plot} and \ref{fig:multi-loc-ranks}A). Thus \texttt{JHUAPL-Bucky} achieved a worse
allocation score than \texttt{CU-select} (on this particular week) by allocating excess resources to several states,
such as Pennsylvania and Michigan, rather than to states such as Florida and California where these resources would have
reduced unmet need.  These allocation errors resulted from forecasts failing to consistently
capture the relative resource needs across different states. The \texttt{CU-select} model made some similar errors —
most prominently, over-allocating resources to Ohio — but overall, it more successfully forecasted the resource demands
across different locations in relative terms.

On the other hand, \texttt{CU-select} had worse performance as measured by MWIS. Its forecasts were biased downwards,
and it consistently incurred a large penalty for underprediction (Figure \ref{fig:multi-loc-ranks}B). The
\texttt{JHUAPL-Bucky} model had wider predictive intervals which more often included the observed level of daily
hospital admissions. Its MWIS was therefore less severely penalized by under- or over-predictions of the actual
hospitalization counts.

\begin{knitrout}
\definecolor{shadecolor}{rgb}{0.969, 0.969, 0.969}\color{fgcolor}\begin{figure}
\includegraphics[width=\maxwidth]{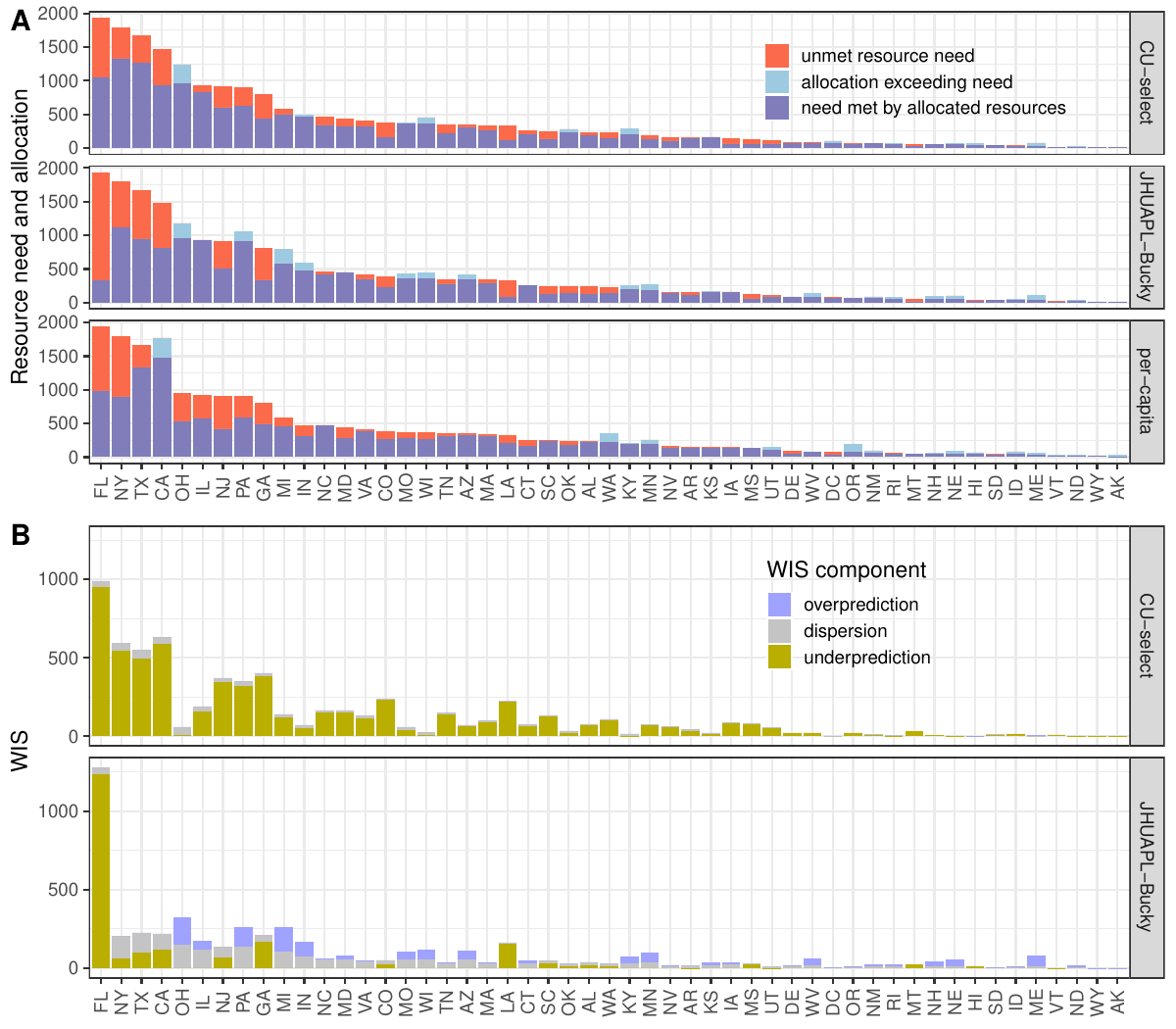} \caption[Component-wise breakdowns of the allocation score (Panel A) and weighted interval score (Panel B), by location for forecasts of hospitalization admissions on January 3, 2022, for two selected models (\texttt{JHUAPL-Bucky} and \texttt{CU-select})]{Component-wise breakdowns of the allocation score (Panel A) and weighted interval score (Panel B), by location for forecasts of hospitalization admissions on January 3, 2022, for two selected models (\texttt{JHUAPL-Bucky} and \texttt{CU-select}). Panel A shows the observed resource need, in this case the observed number of hospitalizations, for each state, along with the hypothetical number of resources allocated to the given location based on the forecasts from each model. The number of available resources was fixed at 15,000 and forecasts from each model were used to determine an optimal allocation strategy before the resource need was known. For most locations the resource need exceeded the resources allocated, indicated by some amount of `unmet resource need' above the `need met by allocated resources'. Panel B shows the breakdown of the weighted interval score (WIS) into components of underprediction, overprediction and dispersion. Larger values of WIS indicate more error, and the full WIS score for each location can be decomposed into the three components shown here.}\label{fig:multi-loc-ranks}
\end{figure}

\end{knitrout}

\subsubsection{Forecast scores showed differences in aggregate and over time}

\begin{knitrout}
\definecolor{shadecolor}{rgb}{0.969, 0.969, 0.969}\color{fgcolor}\begin{figure}
\includegraphics[width=\maxwidth]{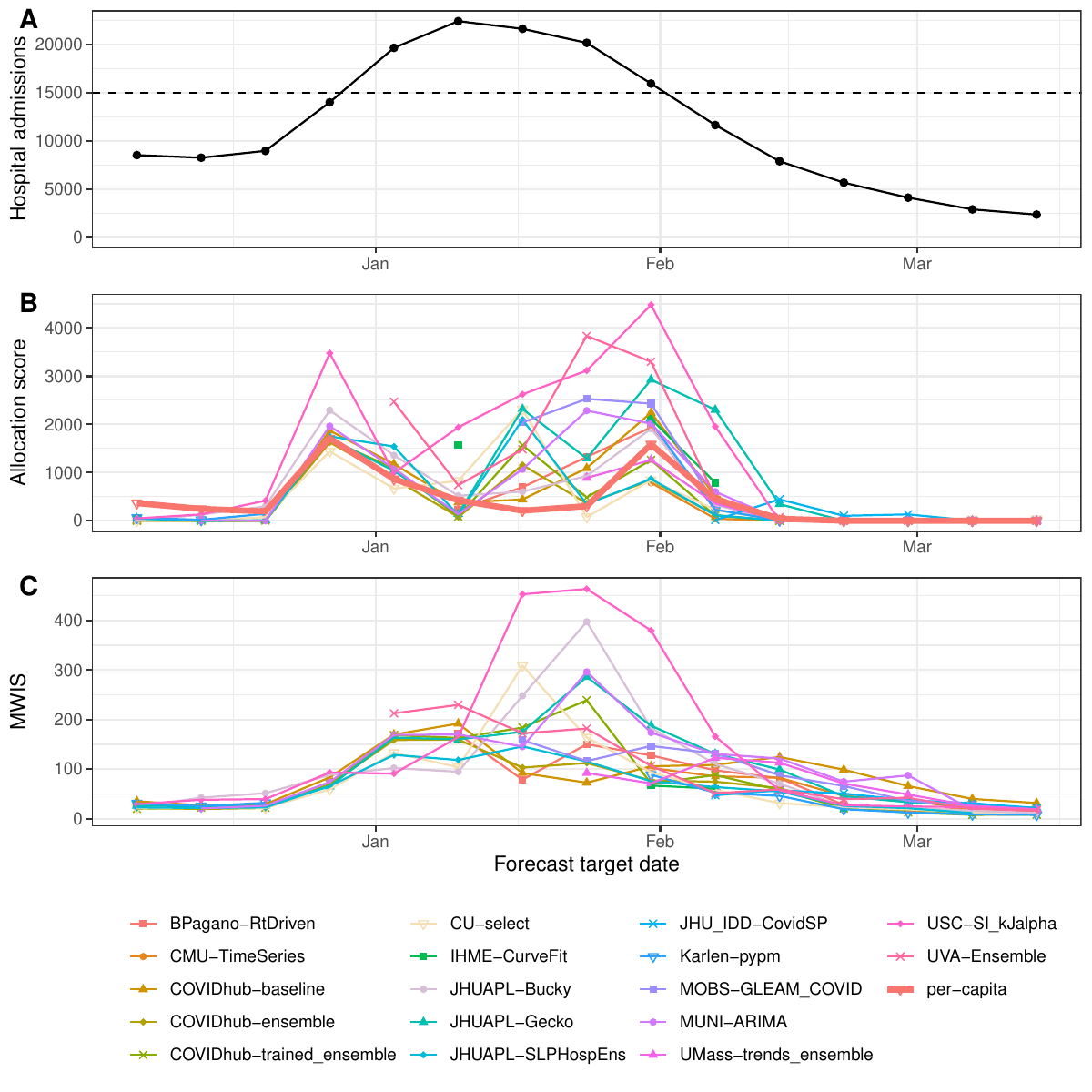} \caption[Hospital admissions and evaluation metrics over time]{Hospital admissions and evaluation metrics over time. Panel A shows the number of hospital admissions in the US as a whole due to COVID-19 on a sequence of 15 Mondays from December 2021 through March 2022. These are the dates for which forecasts were made and evaluated. A horizontal dashed line at 15,000 shows the hypothetical resource constraint $K$. Panel B shows allocation scores (AS) for each model's 14 day-ahead forecast, across all US states. The $x$-axis corresponds to the date of the observation that a model's prediction was targeting (i.e., the date the forecast was made plus the forecast horizon). AS typically are high when the observed value is near to the constraint, which occurs during the last Monday in December (on the way up) and the last Monday in January (on the way down). In Panel B, the \texttt{per-capita} allocation is drawn in a heavier solid line. Panel C shows the MWIS  across weeks. MWIS tends to scale with the observed and predicted values, peaking just after the peak of the Omicron wave.}\label{fig:metrics-over-time}
\end{figure}

\end{knitrout}

Allocation scores varied substantially by date and by model (Figure \ref{fig:metrics-over-time}). For predictions made
for the first three Mondays in December 2021 and the last three Mondays in February 2022 all models had allocation
scores under 500 (and the mean across all models was less than 100), indicating that unnecessary unmet need was fairly
low on those days. The allocation scores were on the whole highest when the observed number of new hospital admissions
was closest to the resource threshold of 15,000.  It was on these occasions that models were most likely to waste resources
in one location that the oracle would have put to use in another.

Averaging across all weeks, the ensemble forecast achieved a better mean allocation score (MAS) than two benchmark
methods. The \texttt{COVIDhub-ensemble} had the best MAS across all weeks, the \texttt{per-capita} allocation had the
second best MAS, and the \texttt{COVIDhub-baseline} model had the sixth best allocation (Table
\ref{tab:multi-week-performance-summary}). No individual model had a better MAS than the per-capita allocation.

Model rankings on mean allocation score (MAS) and time-averaged mean weighted interval score (taMWIS) were broadly
similar, with some places of disagreement. The four most accurate and the five least accurate models were the same
according to both metrics, although not in exactly the same order (Table \ref{tab:multi-week-performance-summary}).
(This comparison excludes the \texttt{per-capita} allocation which does not use forecasts and therefore cannot be
assigned a MWIS.) Notably, the \texttt{JHUAPL-SLPHospEns} model ranked best on taMWIS but had only the fourth best MAS.

% latex table generated in R 4.3.1 by xtable 1.8-4 package
% Fri Mar  1 16:18:55 2024
\begin{table}[ht]
\centering
\begin{tabular}{lrrr}
  \hline
model & MAS & taMWIS & taMWIS rank \\ 
  \hline
\texttt{COVIDhub-ensemble} & 389 & 70 & 2 \\ 
  \texttt{per-capita} & 464 & - & - \\ 
  \texttt{COVIDhub-trained\_ensemble} & 483 & 87 & 4 \\ 
  \texttt{CU-select} & 502 & 81 & 3 \\ 
  \texttt{JHUAPL-SLPHospEns} & 526 & 67 & 1 \\ 
  \texttt{COVIDhub-baseline} & 594 & 93 & 5 \\ 
  \texttt{JHUAPL-Bucky} & 643 & 112 & 7 \\ 
  \texttt{MUNI-ARIMA} & 707 & 116 & 8 \\ 
  \texttt{JHUAPL-Gecko} & 929 & 110 & 6 \\ 
  \texttt{USC-SI\_kJalpha} & 1473 & 155 & 9 \\ 
   \hline
\end{tabular}
\caption{Time averaged mean weighted interval scores (taMWIS) and mean allocation scores (MAS) by model across 13 weeks. These results show the average performance across time for the nine models that submitted forecasts for every week from 2021-11-29 through 2022-02-21, as well
  as for a per-capita allocation. MWIS is not defined for the \texttt{per-capita} allocation which does not use forecasts.} 
\label{tab:multi-week-performance-summary}
\end{table}

\subsubsection{Metrics were not consistently correlated over time}

We can gain some insight into the discordance between MAS and taMWIS by examining the correlation between time-specific
AS and MWIS values of the various models in Figure \ref{fig:metrics-correlation}. We find, for example, no clear
association between the consistent and high MWIS rank of the \texttt{JHUAPL-SLPHospEns} model and its highly variable AS
rank. This contrasts with a clearly positive correlation between MWIS and AS ranks in other models such as
\texttt{Karlen-pypm} and \texttt{USC-SI\_kJalpha}. We also can see in more detail the consistently high performance of the
\texttt{COVIDhub-ensemble} on both scores. And while it did not submit forecasts for enough weeks to be included in
Table \ref{tab:multi-week-performance-summary}, \texttt{CMU-TimeSeries} is an interesting example of a model that
performed consistently well on AS but had only middling MWIS ranks.

\begin{knitrout}
\definecolor{shadecolor}{rgb}{0.969, 0.969, 0.969}\color{fgcolor}\begin{figure}[H]
\includegraphics[width=\maxwidth]{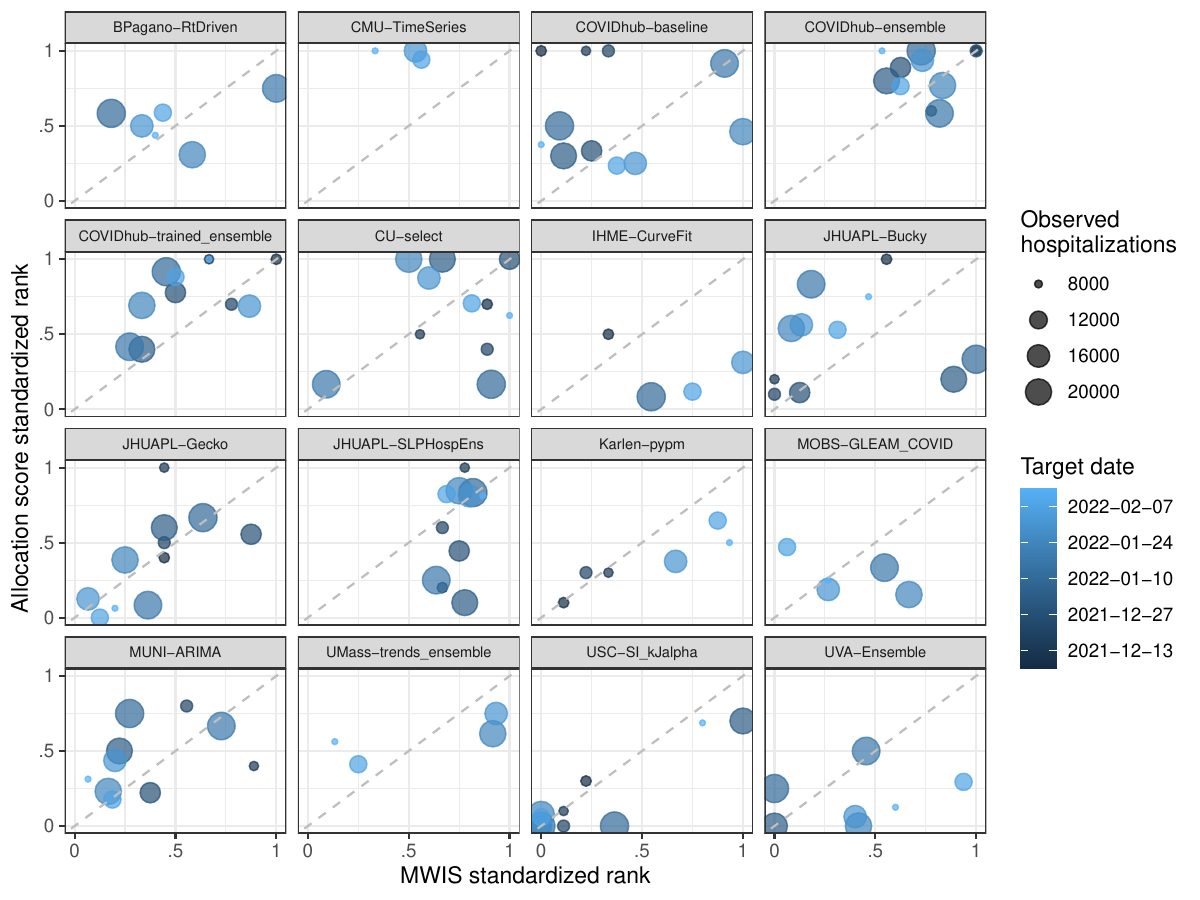} \caption[Association of standardized ranks for MWIS and allocation score by model and week]{Association of standardized ranks for MWIS and allocation score by model and week. Each facet of the plot corresponds to one model. Within each facet, each point corresponds to a week. The $x$- and $y$-values correspond to the MWIS standardized rank and the allocation score standardized rank for that week. Points corresponding to earlier dates have darker shading. The size of the point corresponds to the observed value on the date for which the prediction was made. Models show different degrees of association between the two metrics.}\label{fig:metrics-correlation}
\end{figure}

\end{knitrout}

\subsubsection{Integrated allocation score across values of K}
\label{sec:ias_examp}

AS was computed for a range of $K$ from 200 to 60,000 for forecasts made on December 20, 2020 predicting levels of
hospitalizations on January 3, 2021 as well as for the \texttt{per-capita} allocation (Figure \ref{fig:multi-k}A). These
calculations highlight that AS was highest at at values of $K$ near the observed nationwide total hospital admissions of
19,581 that day. Model ranks by AS were fairly stable across all $K$, and there appears
to be an substantial interval around the observed value over which ranks are constant.

The integrated allocation score (IAS) summarizes allocation scores (AS) across a range of possible values of the
constraint ($K$), allowing one to assign higher weights to more likely values of $K$ (Section
\ref{sec:methods.detailed.integrated_allocation}). IAS was computed for two weight distributions on $K$, one uniform
across the entire range and the other a truncated normal distribution centered at 15,000 (the orange and blue shaded areas in Figure
\ref{fig:multi-k}A, respectively). Both versions of the IAS were correlated with the original AS for $K=15,000$, with
the higher correlation coming from the weighting centered at $K=15,000$ (Figure \ref{fig:multi-k}B). Model rankings
based on the AS and the centered IAS were roughly similar, with the top and bottom three approaches being the same for
both scores (Table \ref{tab:multi-k-scores}).

\begin{knitrout}
\definecolor{shadecolor}{rgb}{0.969, 0.969, 0.969}\color{fgcolor}\begin{figure}[H]
\includegraphics[width=\maxwidth]{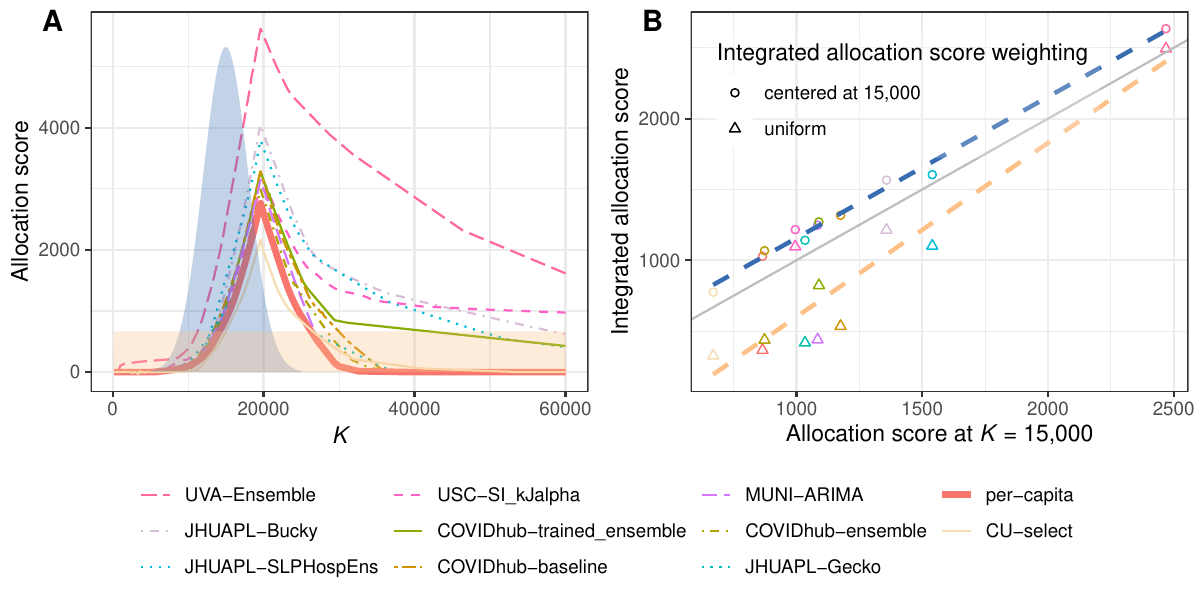} \caption[Allocation scores (AS) across different resource constraints ($K$) for 10 models that made forecasts on 2021-12-20]{Allocation scores (AS) across different resource constraints ($K$) for 10 models that made forecasts on 2021-12-20. Panel A shows, for each model, the AS for values of $K$ between 200 and 60,000 at increments of 200. The AS show a sharp peak just under 20,000, near the eventually observed number of hospitalizations. Two possible weighting functions for the Integrated Allocation Score (IAS) are shown. The first (blue shaded area) computes weights proportional to a normal distribution centered at 15,000 with a standard deviation of 3,000, and truncated to be between 5,000 and 25,000. The second (orange shaded area) uses a uniform weight for all possible values of $K$. Note that the AS used in earlier sections of the application uses the single fixed value of $K=15,000$.  Panel B shows how the two versions of the IAS ($y$-axis) correlate with the AS at $K=15,000$ ($x$-axis). Every point represents the AS at 15,000 and a version of the IAS for one model. As we would expect, the IAS centered at 15,000 (circles) is more closely correlated with the AS at 15,000 than is the uniform IAS (triangles).}\label{fig:multi-k}
\end{figure}

\end{knitrout}

\section{Discussion}
\label{sec:discussion}

Probabilistic forecasts of infectious disease outbreaks have typically been evaluated using well-known proper scoring
rules such as the LS, CRPS, and WIS. Often models are ranked according to such a score, but without reference to any
underlying decision problem from which the score might be derivable as a Bayes scoring rule or be otherwise motivated.
We argue that the value of collaborative forecasting efforts, such as the COVID-19 Forecasting Hub, could be enhanced by
including evaluations of forecasts with
scoring rules that directly consider the success of forecasts in supporting specific public health decisions. As evidence for this
possibility, we have demonstrated how tying forecast evaluation to a specific decision problem (e.g., via the allocation
score) can yield model rankings that differ substantively from those based a current standard measure of forecast
accuracy, the WIS, which does not refer to any decision problem that has been clearly connected to outbreak response.
This result aligns with findings in other fields, especially those where the monetary consequences of using a forecast
can be clearly juxtaposed with standard accuracy measures of the forecast \citep{leitch1991economicForecastEval,
murphy1993whatisagoodforecast, cenesizoglu2012returnPredictionEconValue}. Additionally, we show that an existing
ensemble forecast approach was the only method to outperform (in terms of this new evaluation method) two benchmark
allocation approaches in a hypothetical application, suggesting that there is room for innovation of current
epidemiological modeling techniques that might be spurred on by a reorientation of the field toward practical decision
problems. For example, in order to surpass benchmark allocation scores, forecasters need to devote effort to capturing
the \emph{relative} magnitude of future resource need across different locations.  Success at this task is not, however,
directly rewarded by the WIS. 

Our example application of the AS refers to a hypothetical and unspecified resource need that is assumed to pair with
our forecasted outcome (new hospital admissions) in a manner that might appear overly simplistic and artificially
direct.  We are optimistic, however, that by augmenting this scheme with a probabilistic measurement model for resource
need and more detailed domain knowledge, it could be meaningfully aligned with a more concrete decision problem such as
the allocation of ventilators during a respiratory viral pandemic.  This problem was explored from a stochastic
optimization perspective (not from a forecast evaluation perspective) in \cite{huang_stockpiling_2017} and provided an
initial motivation for this project. We see it as promising focus for further development of the AS framework since it
(a) has been identified as a setting in which allocation decisions are made during respiratory disease outbreaks, (b) involves
logistical considerations depending on geographical and population scales, and (c) would likely imply a central role for
hospitalization data. But again, additional modeling and data synthesis would be required since not everyone who is
hospitalized needs a ventilator. Other possible real-world examples include the allocation of a limited stockpile of
vaccinations \citep{araz_geographic_2012,persad_fair_2023} or diagnostic tests
\citep{du_optimal_2022,pasco_covid-19_2023}.

In practice, epidemiological forecasts are often directed at a diverse set of end-users. It may be easy for some of
these users to summarize the consequences of how they use a forecast with a numerical loss function, but for others,
such as officials deciding how best to update public ``situational awareness'' of an outbreak, this may be essentially
impossible. Even those forecast uses that are easily framed as expected loss minimization may differ enough that no
single score would be appropriate for all users. Ideally, targeted forecasting tools could be developed through close
collaboration between modelers and public health officials. However, this may only be possible in settings with
sufficient staffing on both an analytics and a public health team. Increasingly, collaborative modeling hubs are being
used to generate ``one-size-fits-all'' forecasts for many locations at once. But it could still be beneficial in these
settings to expand the set of scoring methods used by the hub to include scores, such as the AS, which are based on
specific public health decision problems.  This could help end users better understand the value of available forecasts
as inputs to their particular decision making contexts. (We note that issues of propriety are raised when forecasts are
evaluated with scoring rules for which they were not elicited or via parameters imputed by a hub after submission, both
of which are done in our application example.  We consider these issues briefly in appendices
\ref{sec:a:distfromq} and \ref{sec:a:propriety_of_parametric_approximation}.)

There are several important limitations to the current work. The allocation score we developed here does not directly
account for important considerations such as fairness or equity of allocations, or more broadly, individual and group
preference relations that are difficult or even impossible to encode into a loss or utility function.  Ambiguity aversion
(in the sense of \cite{ellsberg1961risk}), for example, seems especially relevant to the use of forecasts in outbreak
response, where there can be immense social and political pressure on public health officials to maintain a transparent
base of evidence for their policy choices and recommendations. The proposed framework also does not attempt to capture
the broader context of decision making. For example, in practice it may be possible to increase the resource constraint
$K$ by shifting funding from other disease mitigation measures. Finally, we emphasize that we opted at this stage of
development to explicitly rule out situations where a successful
epidemiological forecast could lead to policy decisions that change the distribution of the predicted outcome $Y$. Our
framework would need considerable enhancements before being applicable to forecast evaluation beyond horizons for which
causal feedback can be neglected.

An opportunity for further investigation is to more carefully evaluate whether forecasts add value to 
standard  procedures already in place for making public health decisions. In the context of allocations,
such a procedure might extrapolate need from current observed need and population levels (similarly to the two benchmark
approaches presented above), with adjustments based on other political or real-world considerations. For example, in
many settings public health stakeholders will make decisions after synthesizing information from a variety of
quantitative and qualitative sources coupled with expert judgment. The allocation score presented in this work does not
directly measure whether a given forecast adds useful information to such an existing decision-making process. While the
scoring procedures as presented do not directly address this question, they could be modified (say, by comparison to a
baseline model or expert-elicited allocations in the absence of forecast data) to quantify the benefit of using a
forecast to inform a specific decision.

In conclusion, we argue that when possible, the way modelers and policymakers view and evaluate forecasts should more explicitly depend on
the specific decision-making context. Defaulting to standard forecast evaluation metrics can mask the utility (or
disutility) of certain forecasts, or lead to forecasts being used in decision making contexts very different from those
for which they are able to offer useful guidance. New collaborative work between public health officials and modeling teams is
needed to assess the value and relevance of the initial findings presented here, including real-time pilot studies or
simulation exercises that could be used to inform further development of new or alternative scoring metrics. We see this
work as an initial overture for what we hope will grow to be a large, collaborative body of work more closely coupling
applied epidemiological forecasting with public health decision making.

\section*{Acknowledgements}

We wish to thank the following individuals who contributed valuable comments and feedback on early versions of this
work: Matthew Biggerstaff, Rebecca Borchering, Sebastian Funk, Melissa Kerr, and Jeffrey Shaman.

This work has been supported by the National Institutes of General Medical Sciences (R35GM119582) and the U.S.
CDC(1U01IP001122). The content is solely the responsibility of the authors and does not necessarily represent the
official views of NIGMS, the National Institutes of Health, or CDC.

\appendix
\appendixpage
\addappheadtotoc

We address some technical and methodological points from the main text. We begin in section \ref{sec:a:proper} by defining
proper scoring rules and showing that the allocation score is proper. Section \ref{sec:a:bayes-quantiles} gives a
justification for the result that the Bayes act for the allocation problem is given by a vector of quantiles of the
forecast distributions in each location at a shared probability level, which was stated in section
\ref{sec:methods.detailed.specific_allocation}. We describe the algorithm that we use to compute allocations given a
forecast distribution in each location in section \ref{sec:a:numeric}. In section \ref{sec:a:distfromq}, we describe the
methods for approximating a full distribution from a set of quantiles, implemented in the R package \verb`distfromq`,
that we used to support computation of allocation scores from quantile forecasts that were submitted to the US COVID-19
Forecast Hub for the application in section \ref{sec:application}. Finally, in section
\ref{sec:a:propriety_of_parametric_approximation} we examine implications for propriety of an analysis that uses summaries
of forecasts (such as predictive quantiles) rather than the full forecast distributions for computation of allocation
scores.

\section{Proper scoring rules}
\label{sec:a:proper}

In decision theory, a loss function $l$ is used to formalize a decision problem by assigning numerical value $l(x,y)$ to
the \emph{result} of taking an \emph{action} $x$ in preparation for an \emph{outcome} $y$. A \emph{scoring rule} $S$ is
a loss function for a decision problem where the action is a probabilistic forecast $F$ of the outcome $y$ (or the
statement of $F$ by a forecaster). We refer to the realized loss $S(F,y)$ as the \emph{score} of $F$ at $y$.

Probabilistic forecasts can be seen as a unique kind of action in that they can be used to generate their own
(simulated) outcome data, against which they can be scored using $S$. A probabilistic forecast $F$ is thus committed to
the ``self-assessment'' $\Ex_F [S(F, Y)] := \Ex [S(F, Y^F)]$, where $Y^F \sim F$ is the random variable defined by
sampling from $F$, as well to an assessment $\Ex_F [S(G, Y)]$ of any alternative forecast $G$.

A natural consistency criterion for $S$ is that, for observations assumed to be drawn from $F$, it will not assess any
other forecast $G$ as being better than $F$ itself, that is, that
\begin{align}
\Ex_F [S(F, Y)] \leq \Ex_F [S(G, Y)] \label{eqn:prop_ineq}
\end{align}
for any $F,G$. A scoring rule meeting this criterion is called \emph{proper}. If $S$ were improper, then from the
perspective of a forecaster focussed (solely) on expected loss minimization, the decision to state a forecast $G$ other
than the forecast $F$ which they believe describes $Y$ could be superior to the decision to state $F$. $S$ is
\emph{strictly proper} when \eqref{eqn:prop_ineq} is a strict inequality, in which case the \emph{only} optimal decision
for a forecaster seeking to minimize their expected loss is to state the forecast they believe to be true.

\subsection{The allocation score is proper}
\label{sec:a:alloscore_proper}

Our primary decision theoretical procedure, outlined in section \ref{sec:methods.detailed.decisiontheory},
uses a decision problem with loss function $s(x,y)$ to define a scoring rule
\begin{align}
S(F,y) := s(x^F,y) \label{eqn:bayes_sr}
\end{align}
where $x^F := \argmin_{x} \Ex_F[s(x,Y)]$ is the Bayes act for $F$ with respect to $s$.
Such scoring rules, which we call \emph{Bayes scoring rules},
are proper by construction since
\begin{align}
\Ex_F [S(F, Y)] &= \Ex_F [ s(x^F, Y) ] \nonumber \\
 &= \mathrm{min}_{x} \Ex_F [ s(x, Y) ] \quad \text{ (by definition of $x^F$)} \\
 &\leq \Ex_F [ s(x^G, Y) ] \label{eqn:dt_proper_key} \\
 &= \Ex_F [ S(G, Y)]. \nonumber
\end{align}

The allocation scoring rule is Bayes and therefore proper.

We note that in the probabilistic forecasting literature (see e.g., \cite{gneiting2011making}, Theorem 3) what we have
termed Bayes scoring rules typically appear via \eqref{eqn:bayes_sr} where $x^F$ is some given functional of $F$ which
can be shown to be \emph{elicitable}, that is, to be the Bayes act for some loss function $s$. Such a loss function is
said to be a \emph{consistent loss (or scoring) function} for the functional $F \mapsto x^F$, and many important recent
results in the literature (e.g., \cite{fisslerziegel2016consistency}) address whether there \emph{exists} any loss
function that is consistent for $x^F$. Our orientation is different from this insofar as we \emph{begin} by specifying a
decision problem and a loss function of subject matter relevance and use the Bayes act only as a bridge to a proper
scoring rule.  Consistency is never in doubt.

\section{Allocation Bayes acts as vectors of marginal quantiles.}
\label{sec:a:bayes-quantiles}

Here we study the form of the Bayes act for the allocation problem (AP) (equation \eqref{eqn:AP} in section
\ref{sec:methods.detailed.specific_allocation}) of the text:
\begin{align}
    \underset{0 \leq x}{\mathrm{minimize}}\,\, 
    \mathbb{E}_{F} [s_A(x, Y)]= \sum_{i=1}^{N} L \cdot \mathbb{E}_{F_i}[\max(0, Y_i - x_i)]
     \text{ subject to }
     \, \sum_{i=1}^N x_i = K, \label{eqn:AP-long}
\end{align}
where the marginal forecasts $F_i$ for $i=1,\dots,N$ represent forecasts for $N$ distinct locations. We show that the
Bayes act $x^{F,K} = (x_1^{F,K},\ldots,x_N^{F,K})$ for a forecast $F$ and resource constraint level $K$ is a vector of
quantiles of the marginal forecast distributions $F_i$ at a single probability level $\tau^{F,K}$, that is, $x_i^{F,K} =
q_{F_i,\tau^{F,K}}$. An immediate consequence used in the examples in Section \ref {sec:methods.overview} is that if
$F_i = \mathrm{Exp}(1/\sigma_i)$ for all $i$, then the Bayes act is proportional to $(\sigma_1,\ldots,\sigma_N)$, since
$q_{\mathrm{Exp}(1/\sigma),\tau} = -\sigma \log(1-\tau)$.

We begin by noting that a key feature of each term of the loss function $s_A(x, Y)$ defining the AP is the presence of 
a \emph{shortage}: an amount $\max\{0,y-x\}$ by which a resource demand $y$ exceeds a supply decision variable $x$,
which, for convenience, we write as $(y-x)_{+}$.  This is a feature shared with decision problems used to define quantiles 
and related scoring rules such as the
CRPS and the WIS (see e.g., \cite{gneiting2011quantiles}, \cite{jose2009evaluating}, and \cite{royset2022optimization},
sections 1.C and 3.C). In particular, 
a quantile at probability level $\alpha$ of a
distribution $F$ on $\mathbb{R}$ (which we assume to have a well-defined density $f(x)$) is a Bayes act for the loss
function
\[
l(x,y) = Cx + L(y-x)_{+}
\]
where $\alpha = 1-C/L$ and $C$ and $L$ can be interpreted as the cost per unit of a resource (such as medicine) and the
loss incurred when a unit of demand (such as illness) cannot be met due to the shortage $(y-x)_{+}$.  This follows
because a Bayes act, as a minimizer of $\Ex_F[l(x,Y)]$, must also be a vanishing point of the derivative
\begin{align}
\dby{}{x} \Ex_F\left[l(x,Y)\right] &= \Ex_F\left[\dby{}{x}l(x,Y)\right] \nonumber\\
&= C + L\Ex_F\left[\dby{}{x}(Y-x)_+\right] \nonumber\\
&= C - L\Ex_F\left[\mathbf{1}\{Y > x\}\right] \nonumber\\
&= C + L(F(x) - 1), \label{eqn:q_deriv}
\end{align}
so that $1-C/L = F(x)$.
The formula $\dby{}{x}\Ex_F\left[(Y-x)_+\right] = F(x) - 1$ for the derivative of the shortage, used above in
\eqref{eqn:q_deriv}, can be obtained from an application of the ``Leibniz Rule'':
\begin{align}
	\frac{d}{dx} \Ex_F [(Y-x)_{+}] &= \frac{d}{dx} \int_{x}^{\infty} (y-x) f_Y(y)dy \nonumber\\
	&= \int_{x}^{\infty} \frac{d}{dx}(y-x) f_Y(y)dy - (x-x) f_Y(x) = -\int_{x}^{\infty} f_Y(y)dy = F(x)-1. \label{eqn:shortage_deriv}
\end{align}
Note that more care is required when $F$ does not have a density.

Returning to the AP \eqref{eqn:AP-long}, notice that in order for $x^{\star} \in \mathbb{R}^N_{+}$ to be a Bayes act it must be true that reallocating $\delta > 0$ units of
the resource from location $i$ to location $j$ will lead to a net increase in expected shortage --- in other words, the
reallocation increases the expected shortage in location $i$ at least as much as it decreases the expected shortage
in location $j$:
\begin{align*}
&\mathbb{E}_{F_i}[(Y_i - (x^{\star}_i - \delta))_{+}] - \mathbb{E}_{F_i}[(Y_i - x^{\star}_i)_{+}]
\text{ (detriment in $i$) } \nonumber \\
&\qquad \geq
\mathbb{E}_{F_j}[(Y_j - x^{\star}_j)_{+}] - \mathbb{E}_{F_j}[(Y_j - (x^{\star}_j + \delta))_{+}]
\text{ (benefit in $j$) }.
\end{align*}

Dividing by $\delta$ and letting $\delta \searrow 0$, this implies from
\eqref{eqn:shortage_deriv} that
\begin{align}
1-F_i(x^{\star}_i) &= -\frac{d}{dx_i}\mathbb{E}_{F_i}[(Y_i - x^{\star}_i)_{+}] 
\text{ (rate of detriment in $i$) }\nonumber \\
&= \lim_{\delta \searrow 0} \frac{1}{\delta}
\left\{\mathbb{E}_{F_i}[(Y_i - (x^{\star}_i - \delta))_{+}] - \mathbb{E}_{F_i}[(Y_i - x^{\star}_i)_{+}]\right\}
\nonumber \\
&\geq
\lim_{\delta \searrow 0} \frac{1}{\delta}
\left\{\mathbb{E}_{F_j}[(Y_j - x^{\star}_j)_{+}] - \mathbb{E}_{F_j}[(Y_j - (x^{\star}_j + \delta))_{+}]\right\}
 \nonumber \\
 &= -\dby{}{x_j}\mathbb{E}_{F_j}[(Y_j - x^{\star}_j)_{+}] = 1-F_j(x^{\star}_j) 
 \text{ (rate of benefit in $j$).}\label{eqn:ASoptimal1}
\end{align}
(Negative derivatives appear here because our optimality condition addresses how a \emph{decrease} in resources will
 \emph{increase} the expected shortage in $i$ and vice versa in $j$.) 

Since \eqref{eqn:ASoptimal1} also holds with $i$ and $j$
reversed, a number $\lambda$ (a \emph{Lagrange multiplier}) exists such that
$L(1-F_k(x^{\star}_k)) = \lambda$ for all $k \in 1,\ldots,N$.
(We scale by $L$ to facilitate possible future interpretations of $\lambda$ in terms of the partial derivatives
of $\mathbb{E}_{F} [s_A(x, Y)]$.)
That is, $x^{\star}_k$ is a quantile $q_{\tau,F_k}$ for
$\tau = 1 - \lambda/L$. The value of $\tau$ is then determined by the constraint equation
\begin{align}
\sum_{i=1}^N q_{\tau,F_i} = K. \label{eqn:quantiles-sum-to-K}
\end{align}
It is important to note that $\tau$ depends on $F$ and $K$ and is \emph{not} a fixed parameter
of the allocation scoring rule.

\section{Numerical computation of allocation Bayes acts}
\label{sec:a:numeric}

To compute an allocation score $S_A(F,y;K) := s_A(x^{F,K},y)$, we require numerical values for a Bayes act solving the
AP \eqref{eqn:AP-long} --- that is, we must find the specific resource allocations for each location that are determined by
the forecast $F$ under the resource constraint $K$. Assuming we have reliable means of calculating quantiles
$q_{\alpha,F_i}$ of the marginal forecasts $F_i$, these allocations are given by $q_{\tau^{\star},F_i}$ where
$\tau^{\star}$ solves the equation \eqref{eqn:quantiles-sum-to-K}. However, this equation is not analytically tractable
and we must resort to a numerical method for finding an approximation $\tilde{\tau}$ of $\tau^{\star}$.

We have implemented an iterative bisection method that makes use of the fact that $\sum_{i=1}^N q_{\tau,F_i}$ is an
increasing function of $\tau$. The algorithm begins with an initial search interval $[\tau_{L,1}, \tau_{U,1}]$ (such as
$[0,\max_{i}F_i(K)]$) that clearly contains the solution $\tau^{\star}$. At each step $j$ of the algorithm, we evaluate
the total allocation $\sum_{i=1}^N q_{\tau_{M,j},F_i}$ at the midpoint of the search interval,
$\tau_{M,j} = \frac{1}{2}(\tau_{L,j} + \tau_{U,j})$ and continue the search on the narrowed sub-interval
\begin{align}
[\tau_{L,j+1},\tau_{U,j+1}] =
\begin{cases}
[\tau_{L,j}, \tau_{M,j}] & \text{if } \sum_{i=1}^N q_{\tau_{M,j},F_i} \geq K \\
[\tau_{M,j}, \tau_{U,j}] & \text{if } \sum_{i=1}^N q_{\tau_{M,j},F_i} < K.
\end{cases} \nonumber
\end{align}
This search continues until $\tau_{U,j+1} < (1+\varepsilon)\tau_{L,j+1}$ for a suitably small $\varepsilon>0$. We have
implemented this procedure along with the resulting score computations in the R package \verb`alloscore`
\citep{gerding-alloscore} which provided all allocation score values used in the analysis of section
\ref{sec:application}.

Subtleties can arise when the forecast densities $f_i$ vanish or are very small, in which case quantiles are non-unique
or highly variable near a probability level, leading to ambiguity or numerical instabilities{} in the evaluation of
$\sum_{i=1}^N q_{\tau,F_i}$. Additionally, if point masses are present in any of the $F_i$,
\eqref{eqn:quantiles-sum-to-K} will not have a unique solution for some discrete set of constraint levels $K$. We have
adopted conventions for detecting such levels and enforcing consistency in score calculations near them. Through
extensive experimentation, we have determined that these conditions seem to address these challenges with the forecasts
we are working with, but we leave a more rigorous approximation error analysis for later work.

\section{Computing allocations from finite quantile forecast representations}
\label{sec:a:distfromq}

In section \ref{sec:application}, we used the allocation score to evaluate forecasts of COVID-19
hospitalizations that have been submitted to the US COVID-19 Forecast Hub. These forecasts are submitted to the Hub
using a set of 23 quantiles of the forecast distribution at the 23 probability levels in the set $\mathcal{T} = \{0.01,
0.025, 0.05, 0.1, 0.15, \ldots, \allowbreak 0.9, 0.95, 0.975, 0.99\}$, which specify a predictive median and the
endpoints of central $(1 - \alpha) \times 100\%$ prediction intervals at levels $\alpha = 0.02, 0.05, 0.1, 0.2,
\allowbreak 0.3, 0.4, 0.5, 0.6, \allowbreak 0.7, 0.8, 0.9$. For a given week and target date, we use $q_{i,k}$ to denote
the submitted quantiles for location $i$ and probability level $\tau_k \in \mathcal{T}$, $k = 1, \ldots, 23$.

In the event that there is some $k \in \{1, \ldots, 23\}$ for which $\sum_i q_{i,k} = K$, i.e., the provided predictive
quantiles at level $\tau_k$ sum across locations to the resource constraint $K$, the solution to the allocation problem
is given by those quantiles. However, generally this will not be the case; the optimal allocation will typically be at
some probability level $\tau^\star \notin \mathcal{T}$.

To address this situation and support the numerical allocation algorithm outlined in section \ref{sec:a:numeric}, we need
a mechanism to approximate the full cumulative distribution functions $F_i$, $i = 1, \ldots, N$ based on the provided
quantiles. We have developed functionality for this purpose in the \verb`distfromq` package for R \citep{ray-distfromq}.
This functionality represents a distribution as a mixture of discrete and continuous parts, and it works in two steps:
\begin{enumerate}
  \item Identify a discrete component of the distribution consisting of zero or more point masses, and create an
    adjusted set of predictive quantiles for the continuous part of the distribution by subtracting the point mass
    probabilities and rescaling.
  \item For the continuous part of the distribution, different approaches are used on the interior and exterior of the
    provided quantiles:
  \begin{enumerate}
    \item On the interior, a monotonic cubic spline interpolates the adjusted quantiles representing the continuous part
      of the distribution.
    \item A location-scale parametric family is used to extrapolate beyond the provided quantiles. The location and
      scale parameters are estimated separately for the lower and upper tails so as to obtain a tail distribution that
      matches the two most extreme quantiles in each tail. In this work, we use normal distributions for the tails.
  \end{enumerate}
\end{enumerate}
The resulting distributional estimate exactly matches all of the predictive quantiles provided by the forecaster. We use
the cumulative distribution function resulting from this procedure as an input to the allocation score algorithm.

We refer the reader to the \verb`distfromq` documentation for further detail \citep{ray-distfromq}.

\section{Propriety of parametric approximation} % (fold)
\label{sec:a:propriety_of_parametric_approximation}

In practice, open forecasting exercises are generally not able to collect a perfect description of the forecast
distribution $F$ other than in simple settings such as for a categorical variable with a relatively small number of
categories. In settings where the outcome being forecasted is a continuous quantity (such as the proportion of
outpatient doctor visits where the patient has influenza-like illness) or a count (such as influenza hospitalizations),
forecasting exercises have therefore resorted to collecting summaries of a forecast distribution such as bin
probabilities or predictive quantiles. In this section, we address two practical concerns raised by this. First, we
discuss conditions under which it is possible to calculate the allocation score when only summaries of a forecast
distribution are recorded in a submission to a forecast hub. Second, we show that a post hoc attempt to compute the
allocation score based on submitted predictive quantiles may in fact compute an alternative score that is not proper.

\subsection{Propriety when scoring methods are announced prospectively}

We consider a setting where a forecasting exercise (such as a forecast hub) pre-specifies that forecasts will be
represented using a parametric family of forecast distributions $G_\theta(y)$, and the task of the forecaster is to
select a particular parameter value $\theta$. We use $\mathcal{P}$ to denote the collection of all distributions
$G_\theta$ in the given parametric family. For instance, it has recently been proposed that mixture distributions could
be used to represent forecast distributions \citep{wadsworth2023mixture}. Additionally, we note that the functionality
in \verb`distfromq` can be viewed as specifying a parametric family $\mathcal{P}_{\mathrm{dfq}}$ where the parameters
$\theta$ of $G_\theta$ are its quantiles at pre-specified probability levels, and where the shape of any $G_\theta \in
\mathcal{P}_{\mathrm{dfq}}$ over the full range of its support is entirely controlled by these quantiles.

We find it helpful now to formally distinguish between two decision making problems. The first is the public health
decision maker's allocation problem where the task is to select an allocation $x$, with the allocation loss $s_A(x, y) =
\sum_{i=1}^N L \cdot \max(0, y_i - x_i)$ as described in section \ref{sec:methods.detailed.specific_allocation}. The
second is the forecaster's reporting problem where the task is to select parameter values $\theta$ to report. The
forecaster's loss is given by
\begin{align}
s_R(\theta, y) = s_A(x^{G_\theta}, y), \label{eqn:forecaster_theta_loss}
\end{align}
where $x^{G_\theta}$ is the Bayes act for the allocation problem under the distribution $G_\theta$. In words, the loss
associated with reporting $\theta$ is equal to the loss associated with taking the Bayes allocation corresponding to the
distribution $G_\theta$.

Following our usual construction, the Bayes act for the forecast reporting problem is the parameter set that minimizes
the forecaster's expected loss. Breaking with our earlier notation for improved legibility, we use $\theta^\star(F)$ to
denote this Bayes act:
\begin{align*}
\theta^\star(F) &= \text{argmin}_\theta \Ex_F [s_R(\theta, Y)] \\
&= \text{argmin}_\theta \Ex_F [s_A(x^{G_\theta}, Y)]
\end{align*}

We then arrive at the scoring rule
$$S_R(F, y) = s_R(\theta^\star(F), y) = s_A(x^{G_{\theta^\star(F)}}, y).$$
It follows from the discussion in section \ref{sec:a:alloscore_proper} that this is a proper scoring rule for $F$.
Although the full forecast distribution $F$ is not available in the forecast submission, the score $S_R(F, y)$ can be
calculated from the reported parameter values as long as the forecaster submits the optimal parameters
$\theta^\star(F)$.

We emphasize that the forecaster's true predictive distribution $F$ does not need to be a member of the specified
parametric family $\mathcal{P}$ for this construction to yield a proper score. It is, however, necessary to specify the
parametric family to use and the foundational scoring rule $s_A$ (including any relevant problem parameters such as the
resource constraint $K$) in advance, so that forecasters can identify the Bayes act parameter set $\theta^\star(F)$ to
report.

If the parametric family used to represent forecast distributions is flexible enough, the reporting scoring rule $S_R$
and the allocation score will coincide.
Suppose that for a given resource constraint $K$, for any forecast distribution $F$ it is possible to find a member
$G_{\theta^\star}$ of the specified parametric family $\mathcal{P}$ with the same allocation as $F$ (i.e., 
$x^F = x^{G_{\theta^\star}}$). Then $\theta^\star$ is a Bayes act for the reporting problem since for any other
parameter value
$\theta$,
\begin{align*}
\Ex_F[s_R(\theta^\star, Y)] &= \Ex_F[ s_A(x^{G_{\theta^\star}}, Y) ] \\
&= \Ex_F[s_A(x^F, Y)]  \quad \text{ (since $x^F = x^{G_{\theta^\star}}$)} \\
&\leq \Ex_F[ s_A(x^{G_\theta}, Y) ]  \quad \text{ (by definition of $x^F$)} \\
&= \Ex_F[ s_R(\theta, Y)].
\end{align*}
Thus $S_R(F, y) = s_R(\theta^\star, y) 
= s_A(x^{G_{\theta^\star}}, y) 
= s_A(x^F, y) 
= S_A(F, y)$.

For the particular choice of the parametric family $\mathcal{P}_{\text{dfq}}$ (i.e., using the \verb`distfromq`
package), this flexibility requirement is satisfied. For instance, the forecaster could pick one required quantile level
(such as 0.5, for which the corresponding predictions are predictive medians), and set the submitted quantiles of their
forecast distribution in each location at that level to be the desired allocations, which sum to $K$ across all
locations. However, this representation of the forecast may be quite different from the actual forecast distribution
$F$. For example, for the actual forecast distribution $F$ the allocations may occur at some quantile level other than
0.5.

As another alternative for practical forecasting exercises, a forecast hub could ask forecasters to directly provide the
Bayes allocations associated with their forecasts for one or more specified resource constraints $K$. At the cost of
increasing the number of quantities solicited by the forecast hub, this would have several advantages: it would prevent
any artificial distortion of the forecast distributions, allow for direct calculation of scores, and narrow the gap
between model outputs and public health end users. For this to be feasible, implementations of the allocation algorithm
would have to be provided to participating forecasters in the computational languages being used for modeling.

\subsection{Impropriety of post hoc allocation scoring with quantile forecasts}

A \emph{post hoc} evaluation of quantile forecasts that combines the parametric family specified by \verb`distfromq`
with the allocation score does not yield the allocation score of the forecast distribution $F$. Instead, it computes an
alternative score that is improper. This is because the forecast distribution $F$ and the distribution $G^q \in
\mathcal{P}_{\text{dfq}}$ with the same quantiles as $F$ may determine different resource allocations. In our
investigations, these discrepancies appear to be relatively minor on the interior of the provided quantiles, but could
be severe if the tail extrapolations performed by \verb`distfromq` do not match the tail behavior of $F$ and the
allocations are in the tails of the predictive distribution.

We define
$$G^{\star}(F) := \argmin_{G \in \mathcal{P}_{\mathrm{dfq}}} E_{F}[S_A(G, Y)].$$
Since $S_R$ is defined as the Bayes scoring rule for the forecaster's loss \eqref{eqn:forecaster_theta_loss}, $G^{\star}(F)$ coincides
with $G_{\theta^{\star}(F)}$, the distribution in $\mathcal{P}_{\mathrm{dfq}}$ given by the
optimal submission parameters $\theta^{\star}(F)$ for the forecaster with predictive distribution $F$.
In general, $G^q(F)$ and $G^\star(F)$ will be different distributions:  matching $F$ at specific quantiles
does not require $G^q(F)$ to match $F$ at the quantiles for $\tau^{F,K}$ (c.f. \eqref{eqn:quantiles-sum-to-K}),
which would be necessary for it to share $x^F$ as an optimal allocation.

When an analyst attempts a post hoc computation of the allocation score using $G^q(F)$ (implicitly assuming that $G^q(F)
= G^\star(F)$), they in fact compute the alternative score
$$\tilde{S}(F, y) = S_A(G^q(F), y) = s_A(x^{G^q(F)}, y).$$
This score is improper because $E_{F}[S_A(G, Y)]$ is minimized by $G^\star(F)$, not $G^q(F)$.
In general, we have
\begin{align}
E_{F}[\tilde{S}(G^\star(F), Y)] &\leq E_{F}[ S_A(G^q(F), Y) ] \label{eqn:tilde_s_improper} \\
  &= E_{F}[\tilde{S}(F, Y)] \nonumber
\end{align}
However, the inequality in \eqref{eqn:tilde_s_improper} will typically be strict. For example, if $F$ has heavy upper
tails (such as for a lognormal distribution), but normal distributions are used for tail extrapolations in
\verb`distfromq`, then the resource allocations based on the distribution $G^q(F)$ may be quite different from the
optimal allocations under the distribution $F$, leading to a strict inequality. This demonstrates that $\tilde{S}$ is
improper.

\end{document}